\documentclass[12pt,letter]{article}
\pdfoutput=1
\usepackage{graphicx, epsfig, color,cite,inputenc}
\usepackage{amsmath}
\usepackage{amssymb}
\usepackage{float}
\usepackage{caption,subcaption,graphicx}
\usepackage{hyperref}
\usepackage{slashed}
\usepackage[compat=1.1.0]{tikz-feynman}

\def\to{\rightarrow}

\def\bi{\begin{itemize}}
\def\ei{\end{itemize}}

\def\tf{\tilde f}

\def\tg{\tilde g}

\def\alt{\lesssim}
\def\agt{\gtrsim}
\def\be{\begin{equation}}  
\def\ee{\end{equation}}  
\def\bea{\begin{eqnarray}}  
\def\eea{\end{eqnarray}}

\begin{document}
\begin{titlepage}
\begin{flushright}
OU-HEP-220401
\end{flushright}

\vspace{0.5cm}
\begin{center}
{\Large \bf On dark radiation from string moduli decay to ALPs
}\\ 
\vspace{1.2cm} \renewcommand{\thefootnote}{\fnsymbol{footnote}}
{\large Howard Baer$^{1}$\footnote[1]{Email: baer@ou.edu },
Vernon Barger$^2$\footnote[2]{Email: barger@pheno.wisc.edu} and
Robert Wiley Deal$^{1}$\footnote[3]{Email: rwileydeal@ou.edu} 
}\\ 
\vspace{1.2cm} \renewcommand{\thefootnote}{\arabic{footnote}}
{\it 
$^1$Homer L. Dodge Department of Physics and Astronomy,
University of Oklahoma, Norman, OK 73019, USA \\[3pt]
}
{\it 
$^2$Department of Physics,
University of Wisconsin, Madison, WI 53706 USA \\[3pt]
}

\end{center}

\vspace{0.5cm}
\begin{abstract}
\noindent
We examine the issue of dark radiation (DR) from string moduli decay into
axion-like particles (ALPs). In KKLT-type models of moduli stabilization,
the axionlike phases of moduli fields are expected to decouple whilst in
LVS-type moduli stabilization some can remain light and may constitute
dark radiation.
We evaluate modulus decay to Minimal Supersymmetric Standard Model (MSSM)
particles and dark radiation for more general compactifications.
In spite of tightening error bars on $\Delta N_{eff}$, we find only mild
constraints on modulus-ALP couplings due to the somewhat suppressed
modulus branching fraction to DR owing to the large number of MSSM decay modes.
We anticipate that future CMB experiments with greater precision on $\Delta N_{eff}$
may still turn up evidence for DR if the ALP associated with the
lightest modulus field is indeed light.
\end{abstract}
\end{titlepage}

\section{Introduction}
\label{sec:intro}

String theory\cite{Green:1987sp,Green:1987mn} offers a consistent and finite quantum theory of gravity
which can include the matter states and gauge symmetries that form the basis
for the Standard Model (SM). The price to pay is that string theory must be
formulated in 10 (or 11 for $M$-theory) spacetime dimensions.
To gain our observable four spacetime dimensions, the additional space
dimensions are assumed compactified into a tiny compact manifold
such as a Calabi-Yau space (which preserves some spacetime supersymmetry
(SUSY) under compactification).
The resultant 4-d theory then consists minimally of the supersymmetric SM
(MSSM) plus at least an assortment of gravitationally coupled scalar fields
(the moduli) with no classical potential.
The moduli fields, which parametrize the size and shape of the compact space,
must be stabilized and then their vacuum
expectation values (vevs) determine many features of the low energy
4-d effective theory such as coupling constants and soft SUSY breaking terms.
Typical string compactifications then contain, in the 4-d limit,
on order of tens-to-hundreds of
moduli fields in addition to visible and hidden sector fields.

In II-B string theory, the moduli can be classified as complex structure
($U_\alpha$) and K\"ahler ($T_{\beta}$) along with the axio-dilaton ($S$). 
Under {\it flux compactifications}\cite{Douglas:2006es},
the $S$ and $U_{\alpha}$ moduli are stabilized by flux and should gain
ultra-high KK-scale masses.
Under KKLT stabilization\cite{Kachru:2003aw}, the $T_{\beta}$ are
stabilized non-perturbatively while in LVS stabilization\cite{Balasubramanian:2005zx}, the
$T_{\beta}$ are stabilized by a balancing of perturbative and
non-perturbative effects. Thus, the K\"ahler moduli may have much
lighter masses which may be as low as the soft SUSY breaking scale
$m_{soft}\sim 1$ TeV. The lightest of the moduli, $\phi$ (with mass labeled here
as $m_\phi$) may be cosmologically dangerous in that
\begin{enumerate}
\item they may live long enough to decay after the onset of BBN thus
  destroying the successful BBN predictions of light element abundances\cite{Coughlan:1983ci,Banks:1993en,deCarlos:1993wie},
\item they may overproduce neutralino dark matter
  (the moduli-induced LSP problem)\cite{Blinov:2014nla},
\item they may overproduce gravitinos if $m_\phi > 2m_{3/2}$
  (the moduli-induced gravitino problem)\cite{Kohri:2004qu,Nakamura:2006uc,Asaka:2006bv,Endo:2006zj,Dine:2006ii}
  where then the gravitinos could decay
  after onset of BBN or overproduce dark matter and
\item they may decay into relativistic particles such as the
  axion-like particles (ALPs)\cite{Cicoli:2012aq,Higaki:2012ar} which are endemic to 4-d string models\cite{Arvanitaki:2009fg}, thus
  potentially violating limits on dark radiation as parametrized by the
  parameter $N_{eff}$, the effective number of light neutrinos which
  inhabit the cosmic soup.
\end{enumerate}
Under KKLT stabilization, the shift symmetry $T\to T+i\alpha$ is destroyed
and the corresponding ALPs, which comprise the phase fields of the $T_\beta$,
are expected to obtain masses comparable to the corresponding
real components: $m_{\text ALP}\agt m_{soft}$.
However, for LVS stabilization, the shift symmetry can survive,
and the corresponding ALPs end up with small but non-zero masses and
thus may comprise a portion of the measured dark radiation.

Indeed, years ago the measured value of $N_{eff}$ seemed somewhat displaced
from the SM value, motivating great excitement that string remnants may have left a detectable imprint on the cosmic microwave background radiation (CMB).
In recent years, increasingly precise CMB measurements have brought $N_{eff}$
more into accord with SM expectations, so that $\Delta N_{eff}$ provides
now increasingly tight constraints on new physics models which include
dark radiation.

The 2018 Planck analysis of cosmological parameters\cite{Planck:2018vyg}
in relation to CMB measurements is able to fit the amount of dark radiation
at 95\% CL as
\be
N_{eff}=2.99^{+0.34}_{-0.33}\ \ (TT,TE,EE+low E+lensing+BAO)
\ee
based upon joint fits to Planck CMB polarization,
lensing and baryon acoustic oscillations (BAO).
From these fits, and using the Standard Model (SM) value $N_{eff}(SM)=3.046$\cite{deSalas:2016ztq},
we will require that at 95\% CL
\be
\Delta N_{eff}<0.29\ \ \ (95\%\ CL,\ Planck, 2018).
\ee

In this paper, we continue our earlier investigation\cite{Bae:2022okh} of the cosmological
moduli problem (CMP) wherein we computed the various modulus decay rates
into the MSSM particles including all phase space and mixing effects which
are routinely ignored in the literature. Once these are known, then one may compute the modulus decay temperature $T_D\simeq \sqrt{\Gamma_\phi m_P}/(\pi g_*/90)^{1/4}$ and use these to implement BBN
constraints ($T_D>T_{BBN}\sim 3-5$ MeV).
One may also compute the modulus oscillation temperature $T_{osc}$, the modulus-radiation equality temperature $T_e$, and the temperature $T_{3/2}$ of
radiation at the time of gravitino decay. Comparing to the inflaton reheat
temperature $T_R$ and neutralino freeze-out temperature $T_{fo}$, then one
may compute the entropy dilution factor $r=S_f/S_0=T_e/T_D$ and the ultimate
non-thermal neutralino abundance and constraints on relic gravitinos.
Assuming a {\it well-motivated} natural SUSY spectrum of MSSM particles,
it was found that very large modulus masses $m_{\phi}\agt 5000$ TeV were
needed (assuming an initial modulus field strength $\phi_0\sim m_P$) to
avoid the moduli-induced LSP problem. Also,
typically $m_\phi<2m_{3/2}$ was needed to avoid the  moduli-induced gravitino
problem. For the well-motivated gravity-mediated SUSY breaking model,
wherein the soft SUSY breaking scale $m_{soft}\sim m_{3/2}$, then the high mass
modulus solution to the CMP would bring physics into conflict with
naturalness of SUSY models which requires sparticles (save light higgsinos)
typically in the several TeV range\cite{Bae:2022okh}\footnote{For natural SUSY models,
  we require models to have low {\it electroweak} finetuning
  with $\Delta_{EW}\alt 30$\cite{Baer:2012up,Baer:2012cf}.}.
An alternative solution which allows
much lighter values of $m_\phi\sim 30$ TeV is to find an anthropic selection
on modulus field strength $\phi_0\sim 10^{-7}m_P$ which allows for a
more comparable dark-matter-to-baryonic-matter ratio $\sim 1-10$\cite{Baer:2021zbj}.
Alternatively, $m_{\phi}\sim m_{weak}$ with $\phi_0\sim m_P$ leads to a
dark matter dominated universe wherein baryons might only occur as diffuse
hardly gravitating clouds and minimal structure for baryonic
matter\cite{Wilczek:2004cr,Tegmark:2005dy,Freivogel:2008qc}.

In the present work, we extend our previous analyses to include the effects
of modulus decay into ALPs, which may occur for LVS moduli stabilization, but
may occur more generally in other (possibly unthought of) 4-d string models.
This addresses a fourth facet of the CMP: the moduli-induced ALP
problem\cite{Cicoli:2012aq,Higaki:2012ar,Higaki:2013lra}.
But first, we review several related works that precede our
contribution, and explain the new aspects of our own work.
Then in Sec. \ref{sec:decay}, we explain our calculation of modulus field
coupling to ALPs and decay rate into dark radiation. In Sec. \ref{sec:Neff},
we present details of our calculation of $\Delta N_{eff}$ in the
{\it sudden decay approximation} and in Sec. \ref{sec:results} we present
numerical results which mainly restrict the values of $m_\phi$
and modulus-ALP coupling $\lambda_{\text ALP}$.
We find that in spite of the tightening error bars from the experimental
determination of $\Delta N_{eff}$, the moduli-induced ALP problem is
perhaps not overly constraining, even for values of $\lambda_{\text ALP}\sim 1$,
unless $m_\phi$ is very large $m_{\phi}\agt 10^4$ TeV wherein the modulus
field begins to oscillate before the end of inflation (where $T_{osc}\agt T_R$, assuming a reheat temperature of $T_R \sim 10^{12}$ GeV).
On the other hand, the anthropic solution to the moduli-induced LSP problem
does not help much with the modulus-induced ALP problem since
decreasing the field strength $\phi_0$ hardly affects the $\phi\to ALPs$
branching fraction.
A brief summary and conclusions are given in Sec. \ref{sec:conclude}.

\subsection{Some previous work}

Consequences of moduli stabilization for the QCD axion were addressed by
Conlon in Ref. \cite{Conlon:2006tq}.
The issue of dark radiation in string models was addressed by Cicoli, Conlon and Quevedo\cite{Cicoli:2012aq} and Higaki and Takahashi\cite{Higaki:2012ar} in 2012 in the context of
LVS, the moduli stabilization scheme wherein the K\"ahler moduli
shift symmetry $T_{\alpha}\to T_{\alpha}+i\alpha$ is maintained under
perturbative stabilization resulting in (nearly) massless ALP partners
$a_{\alpha}$ of the fields $T_{\alpha}=\phi_{\alpha} +ia_{\alpha}$. These papers
pointed out that dark radiation should be generic in string models with
LVS stabilization whilst models with non-perturbative K\"ahler moduli
stabilization should yield $m_{a_{\alpha}}\sim m_{\phi_{\alpha}}$ so that these
models typically do not produce DR from ALPs. Much of this work was inspired
by CMB measurements which at the time seem to favor enhanced
$\Delta N_{eff}\gg 0$. Thus, other papers examined different dark radiation
sources from {\it e.g.} saxion decay to QCD axions $a$ in SUSY axion
models\cite{Graf:2012hb,Hasenkamp:2012ii, Bae:2013qr,Graf:2013xpe}.
In Ref. \cite{Conlon:2013isa}, Conlon and Marsh investigated the effect
of relic ALPs on BBN and on dark matter production rates (axiogenesis). 
In Ref. \cite{Higaki:2013lra}, Higaki {\it et al.} illustrated the severity of the
moduli-induced axion problem as a fourth aspect of the CMP and emphasized
two solutions: decrease the partial width $\Gamma_\phi (\phi\to ALPs)$
or increase the width $\Gamma_\phi (\phi \to$ visible sector particles).
In Ref. \cite{Cicoli:2013ana}, Cicoli presented a minireview on ALPs from string
compactifications, especially for sequestered LVS models.
In Ref. \cite{Allahverdi:2013noa}, Allahverdi {\it et al.} examined
non-thermal dark matter production in sequestered LVS models including dark
radiation; in \cite{Allahverdi:2014ppa}, they pointed out correlations
between dark matter and dark radiation production in sequestered LVS models.
In Ref. \cite{Angus:2014bia}, Angus noted that dark radiation bounds can
effectively rule out certain extended LVS models where the bulk volume is
stabilized by two rather than one moduli fields.
In Hebecker {\it et al.} Ref. \cite{Hebecker:2014gka},
dark radiation predictions from general LVS scenarios are examined.
In Ref. \cite{Cicoli:2015bpq}, a
general analysis of dark radiation in sequestered string models is made: 
by including  addition modulus visible sector decay modes, the DR is significantly reduced.
Acharya and Pongkitvanichkul\cite{Acharya:2015zfk} considered general string
compactifications giving rise to an {\it axiverse}\cite{Arvanitaki:2009fg} and asked
the question: given the general expectation of an axiverse from string
compactifications, then why is $\Delta N_{eff}$ so small?
Supersymmetric axions, dark radiation and inflation were examined in
Ref. \cite{Queiroz:2014ara}.
Takahashi and Yamada\cite{Takahashi:2019ypv} consider an anthropic bound on $\Delta N_{eff}$ in that if
$\Delta N_{eff}$ is too big, then it suppresses the growth of matter
fluctuations in the early universe and hence suppresses structure formation.
In Ref. \cite{Acharya:2019pas}, Acharya {\it et al.} consider the case of multiple light
string moduli and find that the early-on produced dark radiation is significantly diluted but generically WIMP dark matter is overproduced.
They present a scenario where the WIMP DM abundance is reduced due to
annihilation to dark radiation while the DR abundance is reduced by entropy dilution as long as the DM annihilation is prompt.
In Ref. \cite{Reig:2021ipa}, M. Reig considers a stochastic axiverse wherein a
low scale of inflation extending to $T\sim 100$ eV dilutes all relics while
relic axions can still be produced via a maximal misalignment mechanism.
In Ref. \cite{Anchordoqui:2022gmw}, decay of multiple dark matter particles
to dark radiation in different epochs is shown to not alleviate the
tension in the Hubble constant determination.
As we were completing this work, a comprehensive overview of axions
in string theory with implications for dark radiation and inflationary
models appeared in Ref. \cite{Hebecker:2022fcx}.

\section{Modulus decay to ALPs}
\label{sec:decay}

We begin discussion of modulus decays into ALPs by briefly reviewing how the interaction arises within the K\"{a}hler potential.
After describing how this arises in concrete moduli stabilization scenarios, we then illustrate our ``stabilization-agnostic'' approach.

The K\"{a}hler potential for the lightest geometrical moduli typically takes the approximate form
\begin{align}\label{eq:modulusKahlerPotential}
    K
    &\supset
    -n_i
    \log(
        T_i
        +
        \overline{T}_i
    )
\end{align}
where $n_i$ is determined by the form of the compactification manifold volume, $\mathcal{V}$.
In minimal LVS models, the volume is assumed to be of the form
$\mathcal{V} \propto \tau_b^{3/2}$ (where $\tau_b$ is the ``big'' volume modulus)
so that $n \simeq 3$, which gives the required ``no-scale structure''\cite{Witten:1985xb,Nanopoulos:1994as}.
The volume may however take more complicated forms depending on the compactification details.
In {\it e.g.} fibred LVS models such as those considered in \cite{Angus:2014bia}, the volume of the compactification manifold instead goes as $\mathcal{V} \propto \sqrt{\tau_1} \tau_2$ - corresponding to a K\"{a}hler potential $K \supset -\log( T_1 + \overline{T}_1 ) - 2 \log ( T_2 + \overline{T}_2 )$ for the light geometrical moduli.

The K\"{a}hler metric associated to Eq. \ref{eq:modulusKahlerPotential} is diagonal with entries $K_{i i} \simeq \frac{n_i}{4\tau_i^2}$, so that the (non-canonical) kinetic term in the Lagrangian becomes
\begin{align}
    \mathcal{L}
    &\supset
    K_{i \overline{\jmath}}
    \partial T^i
    \partial \overline{T}^{ \overline{ \jmath } }
    =
    \sum 
    \limits_i
    \frac{n_i}{4 \tau_i^2}
    \partial_\mu
    \tau_i \,
    \partial^\mu
    \tau_i
    +
    \frac{n_i}{4 \tau_i^2}
    \partial_\mu
    c_i \,
    \partial^\mu
    c_i
\end{align}
where the $T_i=\tau_i+ic_i$.
Through the field redefinitions
\begin{align}
    \tau_i
    &=
    \exp 
    \left(
        \sqrt{
            \frac{2}{n_i}
        }
        \phi_i
    \right)
    \quad 
    \text{and}
    \quad 
    c_i
    =
    \sqrt{
        \frac{2}{n_i}
    }
    a_i
\end{align}
the canonical kinetic terms are recovered and after expanding the exponential, we obtain the interaction terms:
\begin{align}\label{eq:interactionGeometricBasis}
    \mathcal{L}
    &\supset
    -
    \frac{1}{m_P}
    \sqrt{
        \frac{2}{n_i}
    }
    \phi_i \,
    \partial_\mu
    a_i \,
    \partial^\mu
    a_i
    +
    \text{higher order}
\end{align}
where we have now explicitly restored $m_P$.
The form of this interaction term loosely matches the expectation for a $\phi_i a_i a_i$ interaction where the $a_i$ terms possess a shift symmetry.
The coupling here is then set explicitly by the field space geometry.
However, for models with more than one light modulus, the $\phi_i$ fields may still need to get rotated into the mass eigenbasis.
While the specific form of the mass matrix is determined by the moduli stabilization details and hence is model dependent, the generic feature is that each modulus mass eigenstate may then decay to multiple ALP types, $a_i$, with couplings determined by $n_i$ and the form of the mass matrix.

In this work, we take the general form of Eq. \ref{eq:interactionGeometricBasis} but make some minor adjustments.
We begin by first writing down the interaction term as
\begin{align}\label{eq:interactionEFT}
    \mathcal{L}
    &\supset
    -
    \frac{\lambda_{\rm ALP}}{m_P}
    \phi \,
    \partial_\mu
    a \,
    \partial^\mu
    a .
\end{align}
Here, we assume that the lightest modulus $\phi$ is written in the mass eigenbasis.
Since that may introduce couplings to multiple ALPs $a_i$ (which we expect to be nearly massless if the shift symmetry is preserved and kinematically inaccessible if the symmetry is broken), we choose to parameterize this by a single ALP field $a$ with an effective coupling $\lambda_{\rm ALP}$.
The $\lambda_{\rm ALP}$ coupling then parameterizes the ``total'' coupling between the lightest modulus and the (possibly many) ALPs it may decay into, without relying on a specific stabilization model.
In the familiar case of minimal LVS, we have $\lambda_{\rm ALP} = \sqrt{2/3} \simeq 0.816$.
Larger values of $\lambda_{\rm ALP} \gtrsim 0.816$ can then parameterize decays to multiple ALPs, whereas $\lambda_{\rm ALP} \lesssim 0.816$ may correspond to some (possibly yet undiscovered) stabilization scheme in which the shift symmetry is broken for some ALPs coupled to $\phi$, making those decays kinematically forbidden and hence lowering the effective coupling.

The matrix element squared for this decay, taking for now a small but
non-zero ALP mass, is 
\begin{align}
    |\mathcal{M}|^2
    &=
    \frac{\lambda_{\rm ALP}^2}{m_P^2}
    m_\phi^4 
    \left(
        1 
        -
        2
        \frac{m_{\rm ALP}^2}{m_\phi^2}
    \right)^2 .
\end{align}
The decay width can then be easily computed:
\begin{align}
    \Gamma_{\phi \rightarrow a a}
    &=
    \frac{\lambda_{\rm ALP}^2}{32 \pi}
    \frac{m_\phi^3}{m_P^2}
    \left(
        1 
        -
        2
        \frac{m_{\rm ALP}^2}{m_\phi^2}
    \right)^2
    \lambda^{1/2}
    \left(
        1,
        \frac{m_{\rm ALP}^2}{m_\phi^2},
        \frac{m_{\rm ALP}^2}{m_\phi^2}
    \right) .
\end{align}
Taking the massless limit for the ALPs $a$, the decay width simplifies to:
\begin{align}
    \Gamma_{\phi \rightarrow a a}
    &=
    \frac{\lambda_{\rm ALP}^2}{32 \pi}
    \frac{m_\phi^3}{m_P^2} .
\end{align}
In the case of {\it e.g.} minimal LVS with $\lambda_{\rm ALP} = \sqrt{2/3}$, we recover the results in \cite{Cicoli:2012aq,Higaki:2012ar}.

\section{Estimating $\Delta N_{eff}$}
\label{sec:Neff}

Our task now is to quantify the amount of dark radiation produced in the
early universe via modulus decay.
As these light, relativistic degrees of freedom should contribute to the
total radiation energy density, $\rho_R^{tot}$, it is useful to parameterize
the radiation density as 
\begin{align}
    \rho_{R}^{tot}
    &=
    \left(
        1 
        +
        N_{eff}
        \frac{7}{8}
        \left(
            \frac{T_\nu}{T}            
        \right)^4
    \right)
    \rho_\gamma   .
\end{align}
Here, $T_\nu$ is the neutrino temperature and $T$ is the temperature of the thermal bath, related by $T_\nu =(\frac{4}{11})^{1/3}T$ \cite{Kolb:1990vq}. 
In the SM, $N_{eff}^{SM} = 3.046$ \cite{deSalas:2016ztq}.
Any additional light degrees of freedom then correspond to an increase in $N_{eff}$ from its SM value, $\Delta N_{eff} = N_{eff} - N_{eff}^{SM}$.
We can then estimate $\Delta N_{eff}$ from modulus decay into ALPs by
\begin{align}
    \Delta N_{eff}
    &=
    \frac{\rho_{\text{ALP}}(T)}{\rho_\nu}
    =
    \frac{120}{7\pi^2}
    \left(\frac{11}{4}\right)^{4/3} 
    \frac{\rho_{\text ALP}(T)}{T^4}
\end{align}
where $\rho_\nu$ is the energy density of a single neutrino species so that $\rho_\nu =\frac{7}{8}\frac{\pi^2}{15}T_\nu^4$.

Since the ALPs dilute as radiation, we can relate their energy density between $T_D$ and $T$ (with $T < T_{D}$) as 
\begin{align*}
    \rho_{\text{ALP}} (T)
    &\simeq
    \left(
        \frac{
            g_{*S}(T)^{1/3}
            \, 
            T
        }{
            g_{*S}(T_D)^{1/3}
            \, 
            T_D
        }
    \right)^4
    \rho_{\text{ALP}} (T_D)
\end{align*}
where we assume conservation of entropy density, $s=\frac{2\pi^2}{45}g_{*S}T^3$, between $T_D$ and $T$.

What remains is to calculate $\rho_{\text{ALP}}(T_D)$. 
Here, to gain the overall big picture, we adopt the sudden decay approximation.
Non-sudden effects will be included in forthcoming coupled Boltzmann calculations.
Using conservation of energy density, we can then express the energy density of the ALPs in terms of the modulus:
\begin{align}
    \rho_{\text{ALP}} (T_D) 
    = 
    \mathcal{B}( \phi \rightarrow a a ) 
    \rho_\phi (T_D) /r 
\end{align}
where $r\equiv S_f/S_i=T_e/T_D$ is the entropy dilution
factor\cite{Choi:2008zq,Baer:2011hx}.

As the modulus behaves as non-relativistic matter, to estimate $\rho_\phi(T_D)$ we first need the number density at $T_D$, which is given by 
\begin{align}
    n_\phi (T_D)
    &= 
    \frac{1}{2}
    m_\phi 
    \phi_0^2 
    \left(
        \frac{
            g_{*S}(T_{D})
            T_{D}^3
        }{
            g_{*S}(T_{\text{osc}})
            T_{\text{osc}}^3
        }
    \right) .
\end{align}
 
\begin{figure}[tbh]
\begin{center}
\includegraphics[height=0.5\textheight]{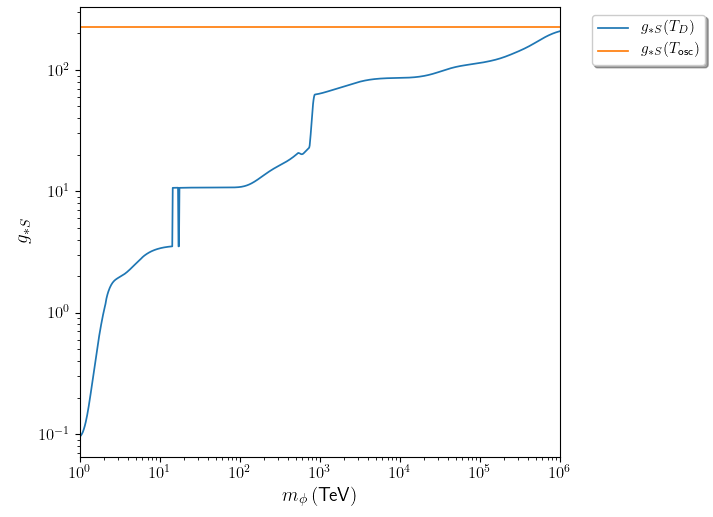}
\caption{The value of $g_{*S}$ vs. $m_{\phi}$ for
  a natural SUSY BM point.
    \label{fig:g*}}
\end{center}
\end{figure}

Putting all of this together, we arrive at the expression for $\Delta N_{eff}$
\begin{align}
    \Delta N_{eff}
    &\simeq
    \frac{60}{7 \pi^2} 
    \left(
        \frac{11}{4}
    \right)^{4/3} 
    \left(
        \frac{ 
            g_{*S}(T)^{4/3}
        }{
            g_{*S}(T_D)^{1/3}
            g_{*S}(T_{\text{osc}})
        }
    \right)
        \mathcal{B}
        \left(
            \phi
            \rightarrow 
            a a
        \right)
    \frac{
        m_\phi^2 
        \phi_0^2 
    }{T_e T_{\text{osc}}^3}
\label{eq:Neff}
\end{align}
where for convenience we show $g_{*S}$ as a function of $T= m_\phi$
in Fig. \ref{fig:g*}.
Furthermore, we note that the temperature of radiation/modulus energy density equality,
determined by requiring $\rho_R=\rho_\phi$, is found to be\cite{Baer:2011hx}
\be
T_e\equiv 
\begin{cases}
    \left(15 / \pi^2 g_*(T_e) \right)^{1/4} \sqrt{m_\phi \phi_0}
    & 
    (T_{osc} < T_{e} < T_R)
    \\
    (3/2 m_P^2)\phi_0^2\sqrt{m_Pm_\phi}
    \left( 10/\pi^2 g_*(T_e) \right)^{1/4}
    &
    (T_e < T_{osc} < T_R)
    \\
    (3/2 m_P^2)\phi_0^2 T_R
    &
    (T_e < T_R < T_{osc} )
\end{cases}
\label{eq:Te}
\ee
and where the oscillation temperature is found to be
\be
T_{osc}\simeq
\begin{cases}
    \left( 10/\pi^2g_*(T_{osc})\right)^{1/4}\sqrt{m_P m_{\phi}}
    &
    (T_{osc} \leq T_R)
    \\
    \left( 10 g_*(T_R)/\pi^2g_*^2(T_{osc})\right)^{1/8}
    \left(
        T_R^2 m_P m_{\phi}
    \right)^{1/4}
    &
    (T_{osc} > T_R),
\end{cases}
\label{eq:Tosc}
\ee
where $m_P\equiv M_{Pl}/\sqrt{8\pi}$ is the reduced Planck mass.

\subsection{Approximate analytic expression for $\Delta N_{eff}$}

We can now make some simple assumptions that should give us a reasonable estimate
for $\Delta N_{eff}$.
Since the leading decay modes for the modulus are to the massless gauge bosons (assuming all couplings as given in \cite{Bae:2022okh} are $\lambda_i \sim 1$),
the total width is well approximated by simply 
\begin{align}
    \Gamma_\phi^{\text{tot}}
    &\sim 
    \Gamma_{\phi \rightarrow \gamma \gamma}
    +
    \Gamma_{\phi \rightarrow g g}
    \sim 
    \frac{9}{8\pi} 
    \lambda_{G}^2
    \frac{m_\phi^3}{m_P^2}
\end{align}
where we make the additional simplifying assumption that $\lambda_{U(1)} \simeq \lambda_{SU(2)} \simeq \lambda_{SU(3)}\equiv \lambda_G$.
The branching ratio can then be approximated as
\begin{align}
    \mathcal{B}
    \left(
        \phi 
        \rightarrow 
        a
        a
    \right)
    &\simeq
    \frac{\lambda_{\text ALP}^2}{36 \lambda_G^2} .
    \label{eq:alpBranching}
\end{align}
giving us the estimate
\begin{align}
    \Delta N_{eff}
    &\simeq
    \frac{5}{21 \pi^2} 
    \left(
        \frac{11}{4}
    \right)^{4/3} 
        \frac{\lambda_{\text ALP}^2}{\lambda_G^2}
    \left(
        \frac{ 
            g_{*S}(T)^{4/3}
        }{
            g_{*S}(T_D)^{1/3}
            g_{*S}(T_{\text{osc}})
        }
    \right)
    \frac{
        m_\phi^2 
        \phi_0^2
    }{T_e T_{\text{osc}}^3}
\end{align}
or, plugging in values
\begin{alignat}{2}
    \Delta N_{eff}
    &\simeq
    0.11
        \frac{\lambda_{\text ALP}^2}{\lambda_G^2}
    && \left(
        \frac{
            224
        }{
            g_{*S}(T_{\text{osc}})
        }
    \right)
    \left(
        \frac{
            g_{*S}(T)
        }{3.9}
    \right)^{4/3}
    \left(
        \frac{63.7}{g_*(T_D)}
    \right)^{1/3}
    \left(
        \frac{m_\phi}{10^6 \text{ GeV}}
    \right)^2
    \nonumber
    \\ &
    && \times 
    \left(
        \frac{\phi_0}{1.94 \times 10^{18} \text{ GeV}}
    \right)^2
    \left(
        \frac{ 
            3.85 \times 10^{11} \text{ GeV}
        }{
            T_{\text{osc}}
        }
    \right)^3
    \left(
        \frac{
            4.0 \times 10^{11} \text{ GeV}
        }{T_e}
    \right) .
\label{eq:Neff_approx}
\end{alignat}

\section{Modulus-to-ALPs branching fraction numerics}
\label{sec:BFs}

In Eq. \ref{eq:Neff}, we found that the amount of dark radiation  depends
directly on the modulus field branching fraction to ALPs.
From Eq.~\ref{eq:alpBranching}, we expect the branching fraction
${\cal B}(\phi\to ALPs)$ to be of
order $\sim 0.01 - 0.1$, although this approximate expression assumes that
the modulus decay to SM modes includes only massless gauge bosons.
The actual result can be very model-dependent in that various MSSM decay
modes may be suppressed or not, and these details can seriously affect the
ultimate modulus branching fraction into DR.

Recently, in Ref. \cite{Bae:2022okh}, all MSSM decay modes of light moduli
fields were computed assuming Moroi-Randall operators\cite{Moroi:1999zb}.
The decay widths were evaluated including all mixing and phase space effects.
Even so, some of these decay modes are still model dependent. In particular,
modulus decay to gauginos may or may not be helicity suppressed.
The helicity suppression is displayed in the decay widths by whether or not the
width numerator contains a factor $m_\phi^3$ (unsuppressed, case {\bf A})
or $m_\phi m_\lambda^2$ (helicity-suppressed, case {\bf B}).
The suppression factor depends on details of the gauge kinetic
  function\cite{Nakamura:2006uc,Dine:2006ii}.
  Likewise,  modulus decay to gravitinos may not be (case {\bf 1})
  or may be  (case {\bf 2}) helicity-suppressed depending on details of the
  K\"ahler function $K$\cite{Nakamura:2006uc,Dine:2006ii}.

  In addition, the moduli branching fractions depend on the specific details
  of the assumed SUSY particle mass spectrum. In Ref. \cite{Bae:2022okh},
  a natural SUSY benchmark (BM) spectrum was adopted with low finetuning
  measure $\Delta_{EW}\sim 20$. The BM point came from the NUHM3 model
  with parameters $m_0(1,2)=10$ TeV, $m_0(3)=5$ TeV, $m_{1/2}=1.2$ TeV,
  $A_0=-8$ TeV and $\tan\beta =10$. It also had $\mu =200$ GeV and $m_A=2$ TeV. 
  Using the Isasugra spectrum calculator\cite{Paige:2003mg}, it is found to
  have $m_h=125.3$ GeV and $m_{\tg}=2.9$ TeV-- in accord with LHC Higgs mass
  measurement and sparticle mass limits. Such a point is expected to emerge
  with a high probability (relative to finetuned models) from string landscape selection\cite{Baer:2022naw,Baer:2022wxe}.

\begin{figure}[tbh]
\begin{center}
\includegraphics[height=0.5\textheight]{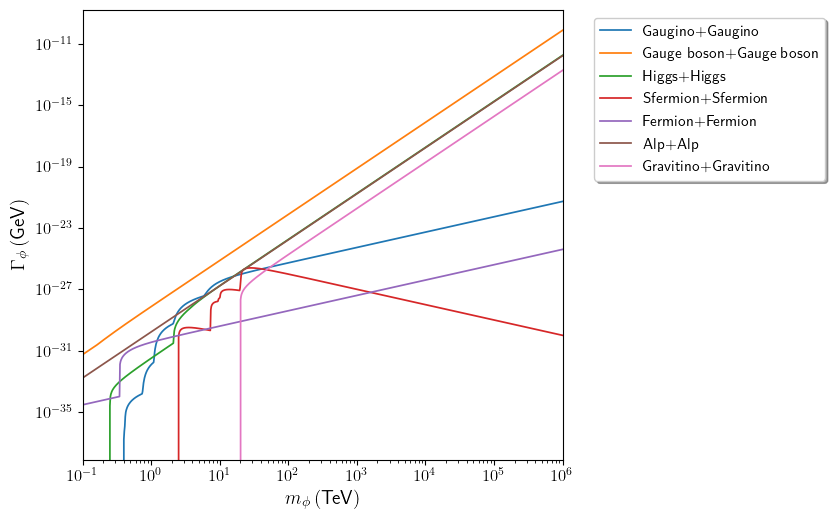}
\caption{Decay widths of lightest modulus field $\phi$
  into ALPs plus various other MSSM particles in the B1 scenario
  with helicity suppressed decay to gauginos but not gravitinos.
  We adopt the natural SUSY BM point from Ref. \cite{Bae:2022okh}.
\label{fig:GammaB1}}
\end{center}
\end{figure}

  In Fig. \ref{fig:GammaB1}, we plot the various modulus to MSSM
  particle partial widths as in Ref. \cite{Bae:2022okh} for the BM point in scenario {\bf B1}, except here
  we also include the partial width for decay to ALPs (brown curve).
  As $m_\phi$ increases, most of the partial widths increase as
  $\Gamma_\phi\sim m_\phi^3$ as commonly assumed in the literature.
  However, here we see the partial widths to gauginos
  (and also SM fermion pairs) increases only as
  $\Gamma_\phi \sim m_\phi^1$ due to the assumed helicity suppression.
    Furthermore, the partial widths into MSSM sfermion pairs suffers additional
    suppression and so
    $\Gamma_\phi (\phi\to sfermions)\sim m_{\tf}^4/(m_\phi m_P^2)$ and thus these partial widths {\it decrease} with increasing $m_\phi$. The important result is that there
      are very many MSSM decay modes, making the ultimate modulus BF into DR
      very model dependent. In much of the literature, it is assumed instead just that $\Gamma_\phi (\phi\to ALPs)\sim \Gamma_\phi (\phi\to MSSM)$.
%

%
%

\begin{figure}[tbh]
\begin{center}
\includegraphics[height=0.5\textheight]{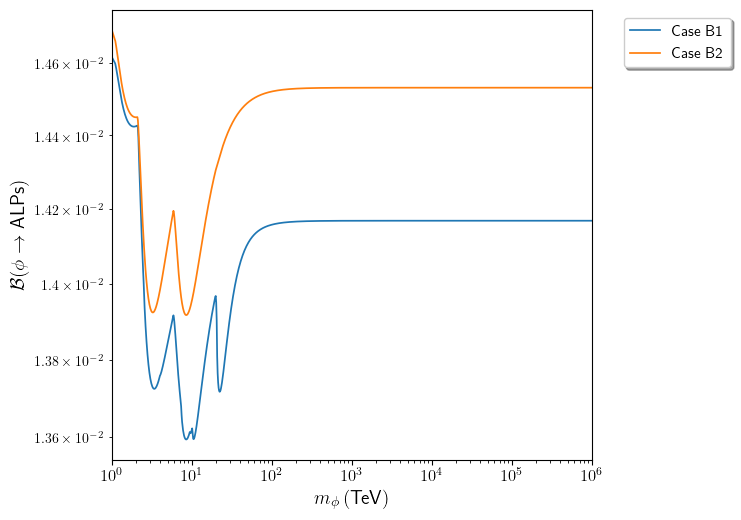}
\caption{Lightest modulus field $\phi$ branching fraction
  into ALPS vs. $m_\phi$ for $\lambda_{\rm ALP}=0.816$ and all other moduli
  couplings $\lambda_i =1$ for the case B1 (blue, case with suppressed decay to
  gauginos but unsuppressed decays to gravitinos) and B2 (orange,
  suppressed decays to gauginos and gravitinos with decays to matter and Higgs superfields turned off)
  with all other MR couplings $\lambda_i =1$ for a natural SUSY BM point from Ref. \cite{Bae:2022okh}.
\label{fig:BFB1B2}}
\end{center}
\end{figure}

In Fig. \ref{fig:BFB1B2}, we show the resultant branching fraction
${\cal B}(\phi\to\ ALPs)$ for the same BM case and scenario {\bf B1}
(blue curve) as in Fig. \ref{fig:GammaB1}.
From the plot, we see that initially at low
$m_\phi$ the BF oscillates somewhat due to turn on of various MSSM
decay modes. Ultimately, the ${\cal B}(\phi\to ALPs)$ settles down to
$\sim 1.4\times 10^{-2}$. The rather low $BF$ result, compared to other
estimates in the literature, is due to our assumption in taking all couplings $\sim 1$,
so that the gauge boson modes (and not the Higgs modes) are dominant - 
whereas in the explicit sequestered models usually studied for DR, 
the gauge bosons typically possess a loop suppression factor.
The inclusion of the very many possible MSSM modes entering the modulus decay width 
then suppresses this branching ratio further to the percent level.

Also in Fig. \ref{fig:BFB1B2}, we show the modulus branching fraction ${\cal B}(\phi\to ALPs)$
versus $m_\phi$ for the same SUSY BM point but in decay scenario {\bf B2}
(orange curve).
This time we assume in addition that the $\phi$ decay modes to Higgs and matter superfields are
turned off, as might be expected if these fields carry Peccei-Quinn (PQ)
charges and are involved in a solution to the strong CP problem
with a Kim-Nilles\cite{Kim:1983dt} solution to the SUSY $\mu$ problem\cite{Bae:2019dgg}.
Although the {\bf B2} scenario is quite different from scenario {\bf B1},
the branching fractions are rather similar since the decays into MSSM
particles are dominated by decays to gauge bosons.

\section{Results for $\Delta N_{eff}$}
\label{sec:results}

\begin{figure}[tbh]
\begin{center}
\includegraphics[height=0.5\textheight]{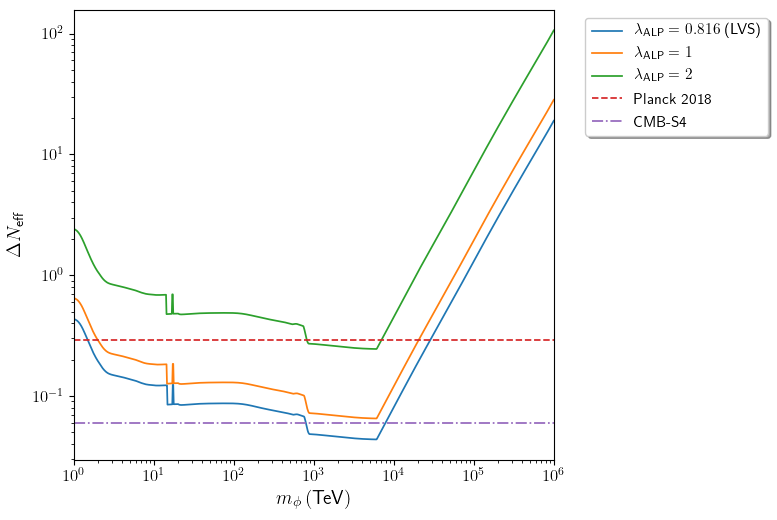}
\caption{The value $\Delta N_{eff}$ vs. $m_{\phi}$ for three cases
  of $\lambda_{\rm ALP}$ along with upper bound from Planck 2018 results
  (red dashed line) and projected reach of CMS-S4 (dot-dashed line).
  \label{fig:Neff}}
\end{center}
\end{figure}

Now that the modulus branching fraction to ALPs is estimated, we can proceed
to using Eq. \ref{eq:Neff} to evaluate $\Delta N_{eff}$ numerically,
including all MSSM decay modes. Our main result is shown in
Fig. \ref{fig:Neff} where we plot $\Delta N_{eff}$ vs. $m_\phi$ for the case
of our SUSY BM point and in scenario {\bf B2}. The horizontal dashed
red line denotes the Planck 2018 95\% CL bound on $\Delta N_{eff}$,
{\it i.e.} one is constrained to live below the red dashed line.
We also show the projected limit of the Stage 4 CMB
experiment\cite{Abazajian:2019tiv} where $\Delta N_{eff}$ as low as
0.06 may be detected (dot-dashed line).
We also show the value of $\Delta N_{eff}$ computed from Eq. \ref{eq:Neff}
assuming three values of $\lambda_{\text ALP}=1,\ 2$ and the minimal LVS value of $0.816$.
From the plot, we see that much of
the green curve for $\lambda_{\rm ALP}=2$ would be excluded,
while the remaining curves are excluded at very low $m_\phi\sim 1-2$ TeV. 
As $m_\phi$ increases,
one might naively expect $\Delta N_{eff}$ to increase as $m_\phi^2$ according
to Eq. \ref{eq:Neff}. But for most of the range, the denominator
$T_eT_{osc}^3\sim m_\phi^2$ as well, so that the $m_\phi$ dependence
roughly cancels out.
Instead, the bulk of $m_\phi$ dependence comes from the $g_*$ degrees of freedom
parameters: namely, as $m_\phi$ increases, then $T_D$ and $T_{osc}$ also change
and consequently the various values of $g_*$ change.
For convenience, we plot in Fig.~\ref{fig:g*} the value of $g_{*S}(T)$ vs.
$T=m_{\phi}$. Here, we see that as various thresholds are passed, then the
effective number of MSSM degrees of freedom can increase significantly
leading to the various slope discontinuities in $\Delta N_{eff}$ of
Fig. \ref{fig:Neff}.

Back to Fig. \ref{fig:Neff}, we see that as $m_\phi$ increases past
$\sim 2$ TeV, then $\Delta N_{eff}$ drops well below the Planck 95\% CL bound for $\lambda_{\text{ALP}} \lesssim 1$
and stays in the allowed region all the way until $m_\phi\agt 20,000-30,000$ TeV whence
$\Delta N_{eff}$ begins a sharp increase into forbidden territory. The sharp
increase is due to the fact that $m_\phi$ is so large that $\phi$ starts
oscillating earlier than our assumed value of reheat temperature
$T_R\sim 10^{12}$ GeV, {\it i.e.} that the field $\phi$ begins oscillating before
the reheating period of inflation has finished. In this case, the dependence of $T_e$ and $T_{osc}$
changes as in Eq's \ref{eq:Te} and \ref{eq:Tosc} and the value of
$\Delta N_{eff}$ consequently increases with increasing $m_\phi$.

Thus, the bulk of the region for $m_\phi\sim 2-30,000$ TeV is allowed by present
CMB data if $\lambda_{\text{ALP}} \lesssim 1$. Furthermore, the $\Delta N_{eff}$ distribution reaches a minimal
value in $\Delta N_{eff}\sim 0.04-0.28$ for $m_\phi\sim 700-6000$ TeV.
This is the same value of $m_\phi$ for which the dark matter abundance
$\Omega_{\chi}h^2$ drops into the allowed region $\alt 0.1$ in Ref. \cite{Bae:2022okh}.
Thus, for these very heavy values of $m_{\phi}$, both the modulus induced
BBN, DM and DR problems are all solved.
A remaining outlier problem would be the moduli-induced gravitino problem
for $m_\phi\agt 2m_{3/2}$. Assuming $m_{3/2}\sim m_{soft}$ as in gravity mediation,
then one is faced with antagonism between SUSY naturalness requiring
TeV-scale soft terms and a solution to the various CMP  problems which may
favor much heavier moduli masses\cite{Bae:2022okh}.
\begin{figure}[tbh]
\begin{center}
\includegraphics[height=0.5\textheight]{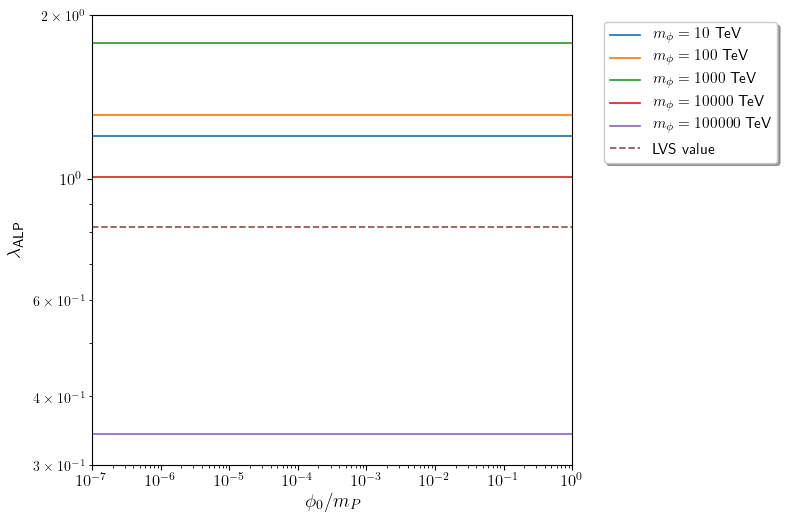}
\caption{Upper bound on $\lambda_{\rm ALP}$ vs. $\phi_0/m_P$ for
  various values of $m_{\phi}$. The dashed line denotes the LVS value of
  $\lambda_{\rm ALP}$.
    \label{fig:lambda_ALP}}
\end{center}
\end{figure}

One way to reconcile the four CMPs with naturalness is to assume the lightest
modulus $\phi$ is stabilized by SUSY breaking so that $m_{\phi}\sim m_{soft}\sim m_{3/2}$ so the gravitino decay mode is closed but with $m_\phi\sim 1-10$ TeV.
The anthropic solution\cite{Baer:2021zbj} to the CMPs then selects
low $\phi_0$ values $\sim 10^{-7}m_P$ in order to gain a comparable
DM-to-baryons abundance.
Notice that the tiny $\phi_0$ value does not suppress the ${\cal B}( \phi\to ALPs)$
since $\phi_0$ tends to cancel in Eq. \ref{eq:Neff}.
To show this more clearly, we plot in Fig.~\ref{fig:lambda_ALP} the
upper bound on $\lambda_{\text ALP}$ versus
$\phi_0/m_P$ which is required to stay below the Planck CMB limit on
$\Delta N_{eff}$ for various values of $m_\phi$.
For most $m_\phi$ values, the upper bound
on $\lambda_{\text ALP}$ exceeds the naive LVS limit although for
$m_\phi\sim 10^5$ TeV the upper limit drops to $\lambda_{\text ALP}\alt 0.31$.
Thus, for much of the range of $m_\phi$, assuming the bulk of $\phi\to MSSM$
decay modes are open, the DR aspect of the CMP does not seem too severe,
at least at present.
On the other hand, at least for the LVS stabilization scheme,
one might expect a good chance for a future deviation in the measured value of
$\Delta N_{eff}$ compared to the SM value.

\subsection{Scatter plots}

\begin{figure}[tbh]
\begin{center}
  \includegraphics[height=0.3\textheight]{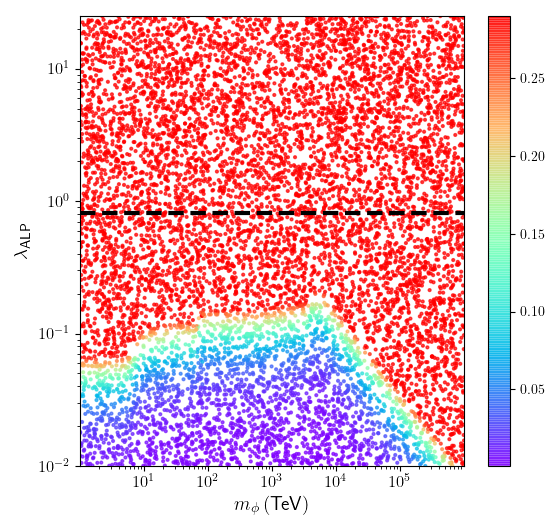}
  \includegraphics[height=0.3\textheight]{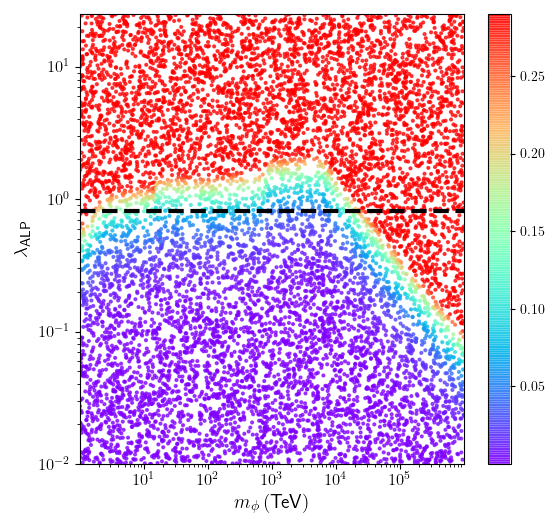}\\
  \includegraphics[height=0.3\textheight]{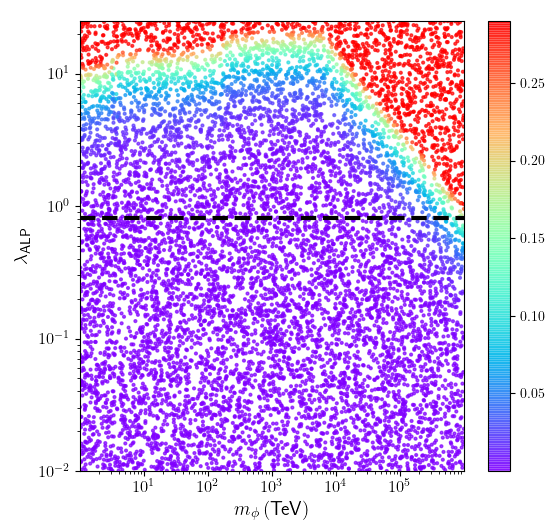}
    \includegraphics[height=0.3\textheight]{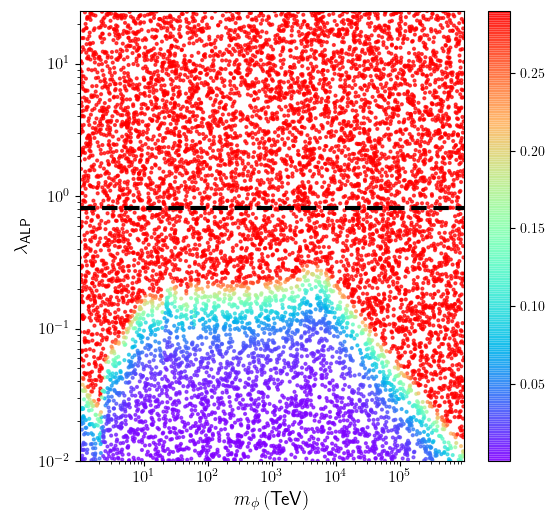}
\caption{Color-coded value of $\Delta N_{eff}$ in the $\lambda_{\text ALP}$ vs.
  $m_\phi$ plane for {\it a}) $\lambda_i=0.1$, {\it b}) $\lambda_i=1$,
  {\it c}) $\lambda_i=10$ and {\it d}) sequestered model with $\lambda_G=0.01$
  but with $\lambda_M=\lambda_H=1$. The dashed line denotes the value of
  $\lambda_{\text ALP}$ in the LVS scenario.
      \label{fig:scatter}}
\end{center}
\end{figure}

To gain more perspective on $\lambda_{\text ALP}$ constraints,
we plot in Fig. \ref{fig:scatter} the allowed
and disallowed regions as scatter plots in the $\lambda_{\text ALP}$ vs. $m_\phi$
plane assuming in the first three frames that all MSSM decay modes
are allowed with values $\lambda_G=\lambda_M=\lambda_H=$ {\it a}) 0.1,
{\it b}) 1 and {\it c}) 10. In frame {\it d}), we adopt the
sequestered LVS model where modulus decays to the gauge sector are suppressed
(with $\lambda_{G}=0.01$) but with modulus decay to Higgs and matter at full strength ($\lambda_M=\lambda_H=1$). The color-coding corresponds to the computed
value of $\Delta N_{eff}$ where red has $\Delta N_{eff}\agt 0.29$ (excluded) and
purple has $\Delta N_{eff}<0.06$ (not even seeable by CMB S4).

For frame {\it a}), with decays to MSSM particles suppressed,
then ${\cal B}(\phi\to aa)$ is large and all $m_\phi$ values with
$\lambda_{\text ALP}\agt 0.1$ are already excluded. As $\lambda_i$ for
$i=G,\ M$ and $H$ increases (thus increasing decays to MSSM particles
and decreasing ${\cal B}(\phi\to aa )$) as in frames {\it b}) $\lambda_i=1$
and {\it c}) $\lambda_i=10$, then more and more of parameter
space becomes allowed (non-red colors). For the sequestered LVS prediction
where decays to gauge bosons and gauginos are suppressed, then again
the ${\cal B}(\phi\to aa)$ becomes large and predicts a value of
$\Delta N_{eff}$  that is beyond Planck 2018 bounds.

\section{Summary and conclusions}
\label{sec:conclude}

In this paper, we have continued our exploration of the various aspects of
the CMP, this time addressing the issue of dark radiation coming from the
lightest modulus $\phi$ decay to its associated ALPs.
In this work, we have neglected a so-called PQ sector QCD axion which may
behave differently from a light stringy ALP in that its interactions would
be suppressed by the PQ scale $f_a$ instead of $m_P$ and its mass and
couplings would be related as in the PQ solution to the strong CP
problem\cite{Baer:2014eja}. For the SUSY PQ model, then one also expects the
presence of saxions and axinos which can all affect the various
particle abundances. In future work, we intend to include light moduli fields
into previous eight-coupled Boltzmann solutions to the dark matter abundance
from mixed axion-neutralino dark
matter\cite{Baer:2011uz,Bae:2013qr,Bae:2014rfa}.

Our present work extends previous work\cite{Baer:2021zbj,Bae:2022okh}
by including modulus field
decay to its associated ALP particle assuming the ALP mass is quite
light as should occur in the LVS moduli stabilization scheme wherein
moduli are stabilized by a combination of perturbative and nonperturbative
effects. In other models such as KKLT with purely non-perturbative K\"ahler
moduli stabilization, then the ALPs tend to get masses comparable to the moduli,
and there is no ALP dark radiation problem.
Our present analysis has been more phenomenological, not restricting
ourselves to a particular stabilization scheme since for more realistic
CY manifolds, one expects far more K\"ahler
moduli than the few that are assumed in simple toy moduli stabilization schemes.
In the more general case, where the full panoply of MSSM decay modes of the
lightest modulus may be allowed and the gauge boson decay modes may be allowed 
at tree level, then the modulus branching fraction to
dark radiation is typically $\sim 0.01$, lower than is usually assumed.
In such cases, where $m_\phi>2m_{\rm ALP}$, then typically the $\Delta N_{eff}$
is below present limits. Future experiments such as Stage 4 CMB
(CMB-S4)\cite{Abazajian:2019tiv} are expected to probe much smaller
$\Delta N_{eff}\sim 0.06$.
If so, then a discrepancy could still
appear for light enough ALP particles coming from string compactifications.

{\it Acknowledgements:} 

This material is based upon work supported by the U.S. Department of Energy, 
Office of Science, Office of High Energy Physics under Award Number DE-SC-0009956 and U.S. Department of Energy (DoE) Grant DE-SC-0017647. 


\bibliography{dr1}
\bibliographystyle{elsarticle-num}

\end{document}